 \documentclass[10pt,preprint]{aastex}

\slugcomment{To appear in 2003 ApJ 586, 464}

\shorttitle{Rotation, magnetism and gyrochronology of solar- and late-type 
  stars}
\shortauthors{Barnes}

\begin{document}


\title{On the rotational evolution of solar- and late-type stars, its 
magnetic origins, and the possibility of stellar gyrochronology\footnote{This
is paper 13 of the WIYN Open Cluster Study (WOCS).}
}


\author{Sydney A. Barnes\altaffilmark{2,3}}
\affil{Astronomy Department, University of Wisconsin-Madison, WI, USA}
\affil{and Astronomy Department, Yale University, New Haven, CT, USA}


\altaffiltext{2}{Visiting Astronomer, Cerro Tololo Inter-American Observatory,
Chile}
\altaffiltext{3}{Visiting Astronomer, Lowell Observatory, Flagstaff, AZ, USA}


\begin{abstract}

We propose a simple interpretation of the rotation period data for solar- and 
late-type stars. The open cluster and Mt. Wilson star observations suggest that
rotating stars lie primarily on two sequences, initially called {\it I} and 
{\it C}. Some stars lie in the intervening gap. 
These sequences, and the fractional numbers of stars on each sequence evolve 
systematically with cluster age, enabling us to 
construct crude rotational isochrones allowing `stellar gyrochronology', 
a procedure, upon improvement, likely to yield ages for individual field stars.

The age and color dependences of the sequences allow the identification of
the underlying mechanism, which appears to be primarily magnetic.
The majority of solar- and late-type stars possess a dominant Sun-like, or 
{\it Interface} magnetic field, which connects the convective envelope both 
to the radiative interior of the star and to the exterior where winds can
drain off angular momentum. These stars spin down Skumanich-style.
An age-decreasing fraction of young G, K, and M\,stars, which are rapid 
rotators, possess only a {\it Convective} field which is not only 
inefficient in depleting angular momentum, but also incapable of coupling 
the surface convection zone to the inner radiative zone, so that only the 
outer zone is spun down, and on an exponential timescale. 
These stars do not yet possess large-scale dynamos. 

The large-scale magnetic field associated with the dynamo, apparently 
created by the shear between the decoupled radiative and convective zones, 
(re)couples the convective and radiative zones and drives a star from the 
Convective to the Interface sequence through the gap on a timescale that 
increases as stellar mass decreases. 
Fully convective stars do not possess such an interface, cannot 
generate an Interface dynamo, and hence can never make such a transition.
Helioseismic results for the present day Sun agree with this scheme, which
also explains the rotational bi-modality observed by Herbst and collaborators 
among pre-main sequence stars, and the termination of this bi-modality 
when stars become fully convective. This paradigm also provides a new basis 
for understanding stellar X-ray and chromospheric activity, light element 
abundances, and perhaps other stellar phenomena that depend on rotation.

\end{abstract}





\section{Introduction} \label{intro}

Beyond a star's brightness and color, and perhaps its metallicity, rotation
is its most fundamental observable characteristic. It is also the one we can
measure to the greatest precision. Rotation periods for fast rotators in open
clusters are routinely measured to one part in ten thousand. This precision,
if it can be harnessed, has great potential to reveal the workings of stars.
But first, it is necessary to understand it. Open cluster stars, whose ages 
are known, allow us to understand rotation as a function of the most 
fundamental, if not generally directly measurable stellar properties, which 
are a star's mass and its age. But stellar rotation has been slow to reveal 
its secrets, and the major steps forward date back to the 1960s and 1970s.
Since then, the situation has become somewhat complicated.

Rotational information for open cluster stars, painstakingly acquired over
the past decades, has revealed a wide, even bewildering, dispersion in the
distribution of stellar rotation rates against spectral type. Because open
clusters are sparse, it has been difficult to identify reliably even the 
existence or absence of any sub-structure in these distributions. 
The fact that the majority of the
rotational information available consists of $v \sin i$ measurements further
complicates this process because of the $\sin i$ ambiguity, and the necessity
of measuring or calculating stellar radii to derive the underlying angular
rotation velocity distribution. Despite these ambiguities, evidence for 
sub-structure in $v \sin i$ data for open clusters has been occasionally 
noted in the literature, especially in relation to the Pleiades 
(eg. Stauffer and Hartman 1987; Soderblom et al. 1993a; Queloz et al. 1998; 
Terndrup et al. 2000).

Stauffer et al. (1984) suggested, and Soderblom et al. (1993b) developed
further the idea that the core-envelope decoupling which 
developed naturally in the rotating stellar models of Endal and Sofia (1981) 
might be related to the observed dispersion in the $v \sin i$ observations. 
This is because angular momentum transport
via hydrodynamic means in the convection zone is very fast, and that in the 
radiative zone is very slow, effectively uncoupling the two regions
(Endal \& Sofia 1981).
However, this model, or the Yale models that inherited that
behavior (Pinsonneault et al. 1989, 1990; Chaboyer et al. 1995; Barnes \& 
Sofia 1996; Sills et al. 2000), or indeed any model that relies on hydrodynamic
means to transport angular momentum within stars decouples, but cannot
re-attach the once decoupled zones, a behavior that contradicts the 
helioseismic observation that the Sun, to zeroth order, is a solid-body 
rotator (Gough 1982, Duvall et al. 1984, Goode et al. 1991, Eff-Darwich et al. 
2002). Thus, the concept of decoupling runs counter to the observed internal
rotational behavior of the Sun.

Since the work of Edwards et al. (1993), it has become the norm to
explain the wide dispersion in rotation rates of cluster stars by invoking
disk interaction on the pre-main sequence, a phenomenon thought to prevent 
some stars from ever passing through a fast rotator phase near the zero 
age main sequence. In this scheme, magnetic interactions between
young stars and their disks create a wide range of initial rotation periods
that converge later on the main sequence.

The difficulty of deciding the internal behavior of stars and the balance 
with the effects of their external
interactions with their disks is further compounded by variations in the 
angular momentum loss rate through stellar winds. In addition to Schatzman's
(1962) original suggestion that only stars with deep surface convection zones
can generate angular momentum loss through magnetized winds, there appears
to be evidence for ``magnetic saturation,'' and hence reduced angular momentum 
loss when stars rotate sufficiently fast (MacGregor \& Brenner 1991), and 
this phenomenon will presumably contribute to some observed features in the 
rotational distributions.

In the observational arena, it has become desirable, and possible, to 
eliminate the ambiguity of $v \sin i$ data by deriving rotation periods 
instead, through photometric monitoring programs aimed at observing light 
modulation by starspots on the surfaces of solar- and late-type stars. 
The demanding nature of these programs and the sparseness of open clusters 
result in small numbers of rotation periods in a given cluster, or indeed, 
for all clusters combined, again making it difficult to identify substructure 
or trends in these rotation period distributions. However, we note
that Van Leeuwen and collaborators, who first discovered the fast rotators in 
the Pleiades (Van Leeuwen \& Alphenaar 1982) and thereby triggered a number of 
subsequent rotational studies, had suggested the existence of ``a pronounced 
gap in the distribution of photometric periods between 0.6 and 6.0 days'' 
(Van Leeuwen et al. 1987). Subsequent observations in other clusters, 
especially the Hyades (Radick et al. 1987) did not immediately reveal 
anything similar.

On the pre-main sequence, 
a notable case of sub-structure in the form of bi-modality has been 
observed in rotation periods for stars in Orion by Attridge \& Herbst (1992), 
and later confirmed by Choi \& Herbst (1996) and Herbst et al. (2001). 
How to connect this observation to rotational observations on the
main sequence is not yet clear, and its interpretation in terms of 
disk-interaction (originally suggested by Edwards et al. 1993) is being 
actively debated in the literature. The situation has become more confusing 
because other researchers deny the presence of bi-modality in similar data 
(eg. Stassun et al. 1999; Rebull 2001), and fail to detect any correlation 
between rotation and disk indicators. 
Furthermore, the rotation rates of the host stars of extra-solar planets 
appear not to be distinguishable from those of a sample of nearby field stars 
(Barnes 2001), as one might expect if disk-interaction is responsible for 
creating slow rotators. These conflicting results are bothersome. 

In a more general sense, the present state of the field of stellar rotation
is highly unsatisfactory. The story of stellar evolution itself requires very 
few basic ingredients and explains much. That of rotational stellar evolution 
invokes many ingredients, some quite arbitrary, and explains little. At the
present time, it is not predictive, and it cannot be used to infer the
rotational history of the Sun in any more than the broadest terms. The 
problem is complicated because of the transitory nature of many of the 
phenomena involved. One needs to understand when a certain phenomenon occurs, 
be it disk-interaction, wind-loss of angular momentum, magnetic saturation, 
de-coupling, internal angular momentum transport etc., and what its importance 
is in the overall scheme. This complexity makes it a field driven largely by 
observations.

In an even more general sense, there are problems in connecting studies of 
young stars in open clusters to those of the mostly older stars in the solar 
vicinity, exemplified by the Mt.\,Wilson stars, which have been individually 
studied in great detail, or for that matter, any sample of nearby field stars. 
One would like to make a solid connection between these studies 
and the incredibly detailed studies of our Sun, in the spirit of the 
solar-stellar connection so that there could be a common basis for 
understanding the apparently disparate results in these fields. 
Finally, in terms of connections to other stellar observables, it is known
that the X-ray or chromospheric emission from stars, their light element
abundances, and a variety of other observables must depend on rotation but
at the present time there is a great degree of confusion about the dependences.

Motivated by these considerations,
we here present a compilation of the entire available rotation period 
database for open clusters on the main sequence, supplemented by recent 
observations in two new clusters, M\,34 and NGC\,3532. 
We also include the Mt.\,Wilson stars, binned into two groups, young and old. 
Despite the age variation among the Mt.\,Wilson stars, this binning makes it 
possible to include them in these rotational considerations to gain the 
leverage in understanding older stars that open clusters do not yet permit.
This scheme lumps together a wide variety of rotational observations of
stars, young and old, in clusters and in the field, and includes several
spectral types. One wonders whether all these observations can be unified
into a single picture.

This paper represents an initial step towards this goal. Here, we describe 
a paradigm for viewing the relevant observational data and place the basic
ideas before the scientific community. In a way, we are using the observations
to decide when the crucial physical phenomena that are involved occur and
what their relative importance is. More complete interpretations 
incorporating stellar models will necessarily follow this work.


\section{Observations and sources} \label{obs}

The principal sources for the cluster rotation periods already available in 
the literature are as follows:
{\sc IC\,2391}: Patten \& Simon (1996); 
{\sc IC\,2602}: Barnes et al. (1999);
{\sc IC\,4665}: Allain et al. (1996);
{\sc Alpha Per}: Prosser \& Grankin (1997), Prosser et al. (1993);
{\sc Pleiades}: Van Leeuwen et al. (1987), Krishnamurthi et al. (1998); 
{\sc Hyades}: Radick et al. (1987); and 
{\sc Coma}: Radick et al. (1990).
We supplement the information available in the literature with 
observations in two open clusters, M\,34 (Barnes et al. 2003, in prep.) and 
NGC\,3532 (Barnes 1998). With ages of approximately 200\,Myr and 300\,Myr 
respectively, these two clusters delineate the phase of rotational evolution 
between the young ($\leq 100$\,Myr) and the Hyades ($\sim 625$\,Myr) open 
clusters.

The M\,34 data were obtained during a 17 night observing run in Sep-Oct 1998 
at the 42'' Hall telescope of Lowell Observatory in Flagstaff, Arizona, 
resulting in a total of 58 visitations to each of two 19'$\times$19' fields in 
M\,34 and $\sim$30 confirmed rotational variables lying on the cluster main
sequence. This field covers only about half the central region of the cluster.
An initial campaign on the other half of the cluster ended in failure. Thus,
this part of the cluster is currently being re-observed, and we envision the
inclusion of all the results, and a complete description of the data analysis
and period derivation in a separate publication. 
The NGC\,3532 data were obtained at the 0.9m telescope at CTIO 
during the course of the author's Ph.D work. This led to the identification
of 87 periodic variables on the cluster sequence, at least 90\% of which may 
be identified with rotational variability of cluster stars. 
The remainder appear to be eclipsing binaries or other unrelated variables. 
This contamination does not affect the conclusions of this paper.

These observations are supplemented by rotation period observations for the
Mt.\,Wilson stars, obtained from Baliunas et al. (1996). The vast majority of
these are observed periods, but a few have been calculated from activity 
indices using the prescription of Noyes et al. (1984). 
The Mt.\,Wilson stars, although not coeval, may be
used for our purposes by classifying them into two groups, young and old, 
as was done originally by Vaughan (1980). Subsequently Soderblom et al. (1991) 
and Donahue (1998) have provided activity ages for these stars. 
Details of this classification, and an appraisal of the rotation periods 
against these ages may be obtained in Barnes (2001). 
The particular result there that is most relevant to this work is that the 
rotation periods of the Mt.\,Wilson stars, when binned by stellar mass, and 
plotted against stellar age, are observed to lie on extremely tight sequences 
that can be very well described by a Skumanich-type spindown (Skumanich 1972). 
The Sun lies precisely on the sequence of the Mt.\,Wilson stars with Sun-like 
masses, at the appropriate location for its age.

Additional observations of higher and lower mass cluster and field stars,
and of pre-main sequence stars will be referred to as necessary. Furthermore, 
it will become obvious that $v \sin i$ data can be used to arrive at the same 
conclusions, albeit with somewhat less certainty.


\section{Morphology of the observations} \label{morph}

The observations are assembled together, ranked by age, and presented 
cluster-by-cluster in the series of panels in Fig.\,1. The measured rotation 
periods are plotted against de-reddened $B-V$ color in panels ranging from an 
age of 30 Myr for the youngest open clusters, IC\,2391 and IC\,2602, to 
$\sim4.5$\,Gyr for the Sun and the old Mt.\,Wilson stars.

The observations display a striking morphology. The rotation periods of open
cluster stars appear not to sample a single population, with a possibly 
wide dispersion that varies with stellar mass and age. Instead, there is 
evidence for sub-populations in these color-period diagrams.
These sub-populations lie on distinct, and relatively narrow sequences in 
these diagrams, and appear to wax and wane with increasing cluster age.
For the moment, we name these the {\it I} and {\it C} sequences, 
and call the cuneiform region between them the {\it gap}\footnote{This 
terminology is chosen because the situation, as we shall see below, 
is reminiscent of the main sequence, giant branch, and the Hertzsprung gap 
in the color-magnitude diagram. We cannot name them the fast and slow sequences
because, as we shall see below, rotation speed alone is not the best descriptor
of these sequences.}.

	\subsection{The {\it I} sequence}

The {\it I} sequence consists of stars which form a diagonal band of 
increasing period with increasing $B-V$ color in a color-period diagram. The 
archetype of this sequence is the sequence of slow rotators in the Hyades open 
cluster. Although it is only faintly visible in the youngest open clusters, 
30-50\,Myr old, accounting for only a small fraction of the total number of 
stars in those clusters, it becomes clearly visible by 200\,Myr (the age of 
M\,34), and accounts for an increasingly large fraction of stars with 
increasing cluster age. All but three stars in the Hyades (625\,Myr) lie on 
this {\it I} sequence.

Of the Mt.\,Wilson stars, whether young or old, all stars redward of $B-V=0.5$ 
lie on the {\it I} sequence. The Sun lies on the {\it I} sequence of the
old Mt.\,Wilson stars\footnote{Note the change in scale for this panel in
Fig.\,1.}. 

This sequence has the feature that it is slowly lifted towards lengthening 
rotation periods as one moves from the youngest open clusters through older 
clusters to the young, and then the old Mt.\,Wilson stars.
Since solar-type stars in the Mt.\,Wilson sample spin down by a factor of $e$
every Gyr or so (see, for example, Fig.\,3 in Barnes 2001), the timescale, 
$\tau_I$, for this spindown is 1\,Gyr. This is the timescale for steady 
angular momentum loss of solar-type stars on the main sequence through 
solar-type winds.

The (positive) slope of the {\it I} sequence increases systematically with
stellar age. Lower mass stars spin down faster. Higher mass stars spin down
slower, or hardly at all. Even among the old Mt.\,Wilson stars, early-type
stars are spinning relatively fast. More massive stars are known not to 
spin down at all (eg. Kraft 1967). This suggests that $\tau_I > 1$ Gyr for 
$M > M_{\odot}$ and $\tau_I < 1$ Gyr for $M < M_{\odot}$.

	\subsection{The radiative sequence}

Blueward of $B-V=0.5$ (late F spectral type), most easily identified with the 
break in the Kraft (1967) curve, and visible also in the Mt.\,Wilson sample, 
is a sequence of higher mass stars which do not have surface convection zones
and do not easily spin down. It appears
to attach smoothly to the {\it I} sequence described above.
For the purposes of this paper, let us call this the {\it radiative sequence}.

	\subsection{The {\it C} sequence}

In young clusters, a sequence of fast rotators, sometimes called the 
ultra-fast rotators (UFRs), is also observed, bifurcating away from 
the {\it I} sequence towards shorter rotation periods. The Pleiades 
observations display this sequence most clearly.
For the present, we call this the {\it C sequence}. 

This sequence is also apparently lifted with time towards lengthening rotation
periods, suggesting spindown of stars on this sequence too. We call this
timescale $\tau_C$. The {\it C} sequence, initially well-populated, becomes 
increasingly sparse with cluster age, boasting only two stars at Hyades age,
and none among either the young or the old Mt.\,Wilson stars.
This sequence bifurcates off the {\it I} sequence. It does not cross 
it. The bifurcation point moves steadily redward from among late F stars for 
the youngest (IC\,2391/IC\,2602) clusters towards K stars for the older, 
Hyades cluster, eventually becoming indiscernable as fewer and fewer stars 
remain on the {\it C} sequence.

The color dependence or slope of this sequence is reversed with respect to 
that of the {\it I} sequence, being either flat for the youngest clusters, 
or displaying shortening rotation period with increasing $B-V$ color for 
older clusters. The observed increasing fraction of stars on the {\it I}
sequence, and the coupled diminishing fraction on the {\it C} sequence imply
that stars evolve steadily from the {\it C} sequence to the {\it I} sequence
on the timescale, $\tau_C$. The motion of the bifurcation point between
the sequences suggests that $\tau_C$ is zero for F-type stars, and increases
towards 300\,Myr for M-type stars. Mid-late M stars apparently spin down on
a very much longer timescale (Delfosse et al. 1998) and we will address
them in a separate section of this paper.

	\subsection{The gap}

The cuneiform region between the {\it C} and {\it I} sequences in a 
given cluster is also often occupied, although more sparsely and with a 
density that decreases with cluster age. 
Unsurprisingly, we call this region the {\it gap}. 
The fraction of stars in this gap (to the total number of stars) is relatively
high at 30-50\,Myr, and steadily decreases, in concert
with the decreasing source population of stars on the {\it C} sequence.

The timescale, $\tau_G$, for stars to move from the {\it C} sequence to the
{\it I} sequence through this gap must be smaller than the timescale, $\tau_C$,
for the {\it C} sequence to approach the {\it I} sequence. In symbols, 
$\tau_G < \tau_C$. The observation that in a given cluster, the number of
stars in this region is lower than the number of stars on the {\it C} 
sequence also corroborates this assertion.

NGC\,3532 offers the clearest display of the two sequences which will dominate 
our discussion, their bifurcation, and the gap. The same data as in Fig.\,1
are again plotted in Fig.\,2, on a logarithmic scale, and with lines
overplotted to guide the eye in locating the sequences.

	\subsection{Migration from one sequence to the other}

Some stars appear to be located on the {\it I} sequence in even the youngest 
open clusters. Those that are not are apparently on the {\it C} sequence, 
or in the region between the sequences. Stars (in this mass range) not 
initially on the {\it I} sequence migrate onto it in less than 1\,Gyr since 
all of the relevant Mt.\,Wilson stars (and presumably the field stars studied 
by Kraft (1967) are on the {\it I} sequence. 

This subjective view is confirmed more quantitatively by the fact that the 
relative fractions of stars on these sequences change systematically with 
cluster age. In the youngest clusters, the fraction of stars on the {\it I}
sequence is roughly one-quarter. This fraction increases steadily with cluster 
age, reaching $90\%$ at Hyades age and $100\%$ among both the young and old 
Mt.\,Wilson samples. The fraction of {\it C} sequence stars diminishes 
steadily 
with age, starting at one-half, and stars in the gap account for the rest. 
This information is illustrated graphically in Fig.\,3, which demonstrates 
that the time-scale for this transition for the unbinned group of G, K, and 
early M stars considered  in this study is $\sim$200-300 Myr. We note that 
these fractions are almost monotonic despite small number statistics and the 
possible selection effects undoubtedly present in rotation period data.


\section{Interpretion} \label{intp}

The foregoing observations can be interpreted in a relatively conventional 
manner using ideas that have already appeared in the literature in one guise 
or another. Here, we merely systematize and weld them into a coherent picture 
by supplying some additional ingredients.
Although winds and angular momentum loss are likely to be considerably more
complicated than we suggest, we begin this process below, laying out only the 
most general ideas here.

\subsection{The {\it I} sequence}

	\subsubsection{The connection between the Sun, Mt.\,Wilson, and open
			cluster stars}

In the observations presented here, the Mt.\,Wilson stars are all observed 
to lie on the {\it I} sequence. The width of this sequence diminishes when we 
account for the individual ages of these stars and bin them accordingly. 
Thus, the observed dispersion of this sequence, relative to that of each
open cluster, must largely be a result of the non-coeval nature of this sample.

When the Mt.\,Wilson stars are binned by stellar mass, then their rotation
periods correlate tightly with their (activity) ages, as was shown recently 
in Barnes (2001). In fact, when binned by mass, the Mt.\,Wilson stars appear 
to obey the Skumanich relationship quite precisely - see especially Fig.\,3 in 
that publication. The Skumanich (1972) relationship is $v \propto t^{-1/2}$, or
equivalently, $P \propto t^{1/2}$, where $t$, $v$ and $P$ are respectively the 
age, rotation velocity, and rotation period of the star. It was originally 
derived in the context of open cluster stars and the Sun. So it offers us a 
means to connect the solar rotation rate and the Mt.\,Wilson observations of 
older stars to those of stars in open clusters, but as we shall see below, 
only to those on the {\it I} sequence. This point of connection between the
Sun, open cluster, and specific groups of field stars has also been noted by 
Soderblom (1983). Before that, Smith (1979) was also impressed by this 
``fortuitously tight relation'' but believed that the Sun was an 
``abnormally rotating star,'' a conclusion we find to be highly unsatisfactory.

In fact, the Sun lies precisely on the sequence delineated by the 
$1\,M_{\odot}$ Mt.\,Wilson stars, at the appropriate location for its age
(see Figs.\,1 \& 2). 
This sequence spins down on a timescale, $\tau_I$, of 1 Gyr for a solar-mass 
star. Higher mass stars spin faster (and spin down slower), and 
lower mass stars spin slower (and spin down faster). 
This behavior is exactly the same as the behavior of stars on the {\it I}
sequence in open clusters, and allows us to identify the two.
Thus, the {\it I} sequence in open clusters spins down Skumanich-style.

	\subsubsection{Rotational isochrones using the {\it I} sequence}

The foregoing facts suggest that it might be possible to account for both the
color dependence and time dependence of rotation to attempt the construction
of rotational isochrones for late-type stars.
In fact, in Fig.\,2, we have overplotted a series of curves, mentioned
earlier as aids to guide the eye, over the {\it I} sequence in each panel. 
The color dependence, apparently separable from the time dependence, is 
arbitrarily chosen here, but can be explained in terms of stellar models later.
The particular functional form we have used here,
incorporating the Skumanich (1972) relationship, is 
\begin{equation}
P = \sqrt{t}\,f(B-V)
\end{equation}
where 
\begin{equation}
f(B-V)=\sqrt{(B-V-0.5)}-0.15(B-V-0.5) 
\end{equation}
and where $P$, $t$, and $B-V$ are the rotation period (d), age (Myr), and 
$B-V$ color respectively. The function, $f$, was arbitrarily chosen to 
represent the color dependence of the data.
These curves form a one-parameter family and that 
parameter is the age, $t$, of the cluster. As a result, these curves together 
represent a crude set of rotational isochrones. 

In order to display the variation more clearly, they are all plotted 
together in the left panel of Fig.\,4, with colors matching the respective
open cluster observations, and widths representing the fraction of cluster
stars on the {\it I} sequence. This method of stellar `gyrochronology'
promises to be especially useful for young open clusters where the usual, 
color-magnitude diagram based techniques, appear to be less reliable.
The relatively narrow widths of the open cluster {\it I} sequences, despite
the doubtless presence of surface differential rotation on stars suggest 
that precisions comparable to conventional techniques are attainable.

Because the Mt. Wilson stars also appear to obey this scheme, we are emboldened
to suggest that the rotation-age relationship might even be true for 
individual field stars. 
The present errors of 30-40\% apparently reflect the inadequacy of the 
relationship provided here, rather than an inherent limitation of the method.

	\subsubsection{Implications for angular momentum loss}

The Skumanich relationship, $v \propto t^{-1/2}$, implies that on the 
main sequence, the loss of angular momentum, $J$, obeys the relationship
$\frac{dJ}{dt} \propto -\omega^3$, where $\omega$ is the star's angular
rotational velocity. The evidence presented here further bolsters the 
veracity of this well-known relationship since it implies and is implied 
by the Skumanich relationship. 
Thus,  $\frac{dJ}{dt} \propto -\omega^3$ must be the appropriate angular 
momentum loss relationship for stars on the {\it I} sequence. 

The origin of this relationship is in magnetized stellar winds, as has been 
shown in prior work (eg. Kawaler 1988), based on earlier work by Mestel (1968) 
and Mestel and Spruit (1987). 
This fact is also well known and has been
used repeatedly to construct models of stellar angular momentum evolution
[eg. Pinsonneault et al. (1989), (1990); Chaboyer et al. (1995); 
Barnes and Sofia (1996); Krishnamurthi et al. (1997); Barnes et al. (2001); 
Sills et al. (2000); Bouvier et al. (1997); Cameron \& Li (1994)]
although usually in a more complicated form originally derived by 
Kawaler (1988) and modified by Chaboyer et al. (1995) to account for the UFRs.

The expression suggested by Kawaler (1988) and used by the
Yale group and many others subsequently is:
  \begin{equation}
  \frac{dJ}{dt}=
           {-K \Omega ^{1+4N/3} (\frac{R}{R_{\odot}})^{2-N}
              (\frac{M}{M_{\odot}})^{-N/3}
              (\frac{\dot{M}}{10^{-14}})^{1-2N/3} 
           }
  \end{equation}
where $K$ is a constant, $\Omega$, $R$, $M$, $\dot{M}$ represent respectively 
the (instantaneous) angular velocity of the star, its radius, its mass, and 
its mass loss rate, and $N$ is a magnetic parameter representing the geometry 
of the magnetic fields, which can vary from $3/7$ to $2$ denoting completely 
dipolar and radial fields respectively. 
$N=1.5$, representing a field that is somewhat more radial than dipolar, 
allowing a mixture of closed field lines and open field lines that escape to 
infinity, reproduces the Skumanich relationship. 
In this case, the Kawaler parameterization above reduces to
  \begin{equation}
  \frac{dJ}{dt}=
           {-K \Omega ^{3} (\frac{R}{R_{\odot}})^{0.5}
              (\frac{M}{M_{\odot}})^{-0.5}
           }
  \end{equation}

Let us re-derive the rotational evolution here to clarify some issues.
The mass, radius, and parameterized magnetic field dependence in this
relationship complicate matters enough that here we prefer to retain
only the simplest part of the above expression, that of the $\omega^3$ 
dependence, so as not to lose the essential thread that connects the ideas.
\begin{equation}
\frac{dJ}{dt} \propto -\omega^3
\Rightarrow \frac{d(I\omega)}{dt} = - C \omega^3
\end{equation}
where $I$ is the appropriate moment of inertia, and $C$ is a constant. Thus, 
\begin{equation}
\frac{d\omega}{\omega^3} = - \frac {C}{I} dt
\end{equation}
assuming that $I$ is constant for a star on the main sequence.
Using the substitution $\omega^2 = s$, we note that on the LHS,
$2 \frac{d\omega}{\omega^3} = \frac{ds}{s^2} = - d(\frac{1}{s})$, so that
\begin{equation}
d(\frac{1}{s}) = 2 \frac {C}{I} dt
\end{equation}
or on integration,
\begin{equation}
\frac{1}{s} - \frac{1}{s_0}= 2C \frac {t}{I}  
\end{equation}
\begin{equation}
\Rightarrow \frac{1}{\omega^2} - \frac{1}{\omega_0^2}= 2C \frac {t}{I} 
\end{equation}
\begin{equation}
\Rightarrow P^2 = P_0^2 + D \frac{t}{I}
\end{equation}
where we have embedded the constants in a new one, $D$.

$P \propto t^{1/2}$ is the original Skumanich relationship for stars of the 
same type, but if we make the very gross assumption that $D$ is the same 
constant for the late-type stars under consideration, then, at a given age, 
the mass dependence of the rotation period $P$ must, to zeroth order, 
be given by: 
  \begin{equation}
	P \propto I^{-1/2} 
  \end{equation}
ignoring the initial period and the effects of stellar evolution, which are
in any case, mild on the main sequence.

	\subsubsection{What is being braked on the {\it I} sequence? }

We are now left with the task of figuring out what $I$ is. The observations
show that $P$ increases with increasing $B-V$ color, so $I$ must decrease.
We note that the {\it I} sequence in several open clusters, although sparse
in some cases at the present time, appears to possess a curvature that suggests
the identification of $I$ with the moment of inertia of the entire star, or a 
fixed fraction thereof. This agrees with the rapid rotation at all ages of 
higher-mass stars and the corresponding slow rotation of their lower-mass 
siblings. It is further bolstered by the long spindown time-scale, $\tau_I$, 
of 1\,Gyr for a solar-mass star.
This suggests that whatever is doing the coupling, which we suggest is the 
magnetic field rather than hydrodynamic forces, couples the whole, or 
substantially the whole, star. It cannot be a hydrodynamic coupling because of 
the contrasting behavior of the {\it C} sequence, discussed later. This 
conclusion is also consistent with the helioseismic observation that the Sun, 
to zeroth order at least, is a solid-body rotator.

Let us continue this simple-minded line of reasoning.
The moment of inertia of a star, $I$, is related to its mass, $M$, and
radius, $R$, via $I = k M R^2$, where $k$ is a structure constant. 
If we make the same gross assumption about homology that we made earlier, 
and since 
$\frac{R}{R_{\odot}} \sim (\frac{M}{M_{\odot}})^{0.7}$ 
on the main sequence, we have
$I \propto (M)^{2.4}$, so that
$P \propto \frac{1}{M^{1.2}}$.
Thus we expect that
\begin{eqnarray}
\lefteqn{P_{1.2 M_{\odot}} : P_{1 M_{\odot}}: P_{0.5 M_{\odot}} }\nonumber \\
   & &  :: 1.2^{-1.2}: 1 : 0.5^{-1.2} 				  \nonumber \\
   & &  :: 0.8       : 1 : 2.3 
\end{eqnarray}
These numbers for the lower-than-solar-mass stars are in rough agreement with 
observations of the {\it I} sequence in several open clusters, the Mt. 
Wilson stars, and the Sun, and suggest that this could indeed be the main 
reason for the shape of the {\it I} sequence in a color-period diagram.

Upon using the moment of inertia of the (300\,Myr-old) stellar models of 
Pinsonneault et al. (1990), one gets 
\begin{eqnarray}
\lefteqn{P_{1.2 M_{\odot}} : P_{1 M_{\odot}}: P_{0.5 M_{\odot}} }\nonumber \\
   & &	:: 10^{-53.95/2}:10^{-53.83/2}:10^{-53.28/2}		 \nonumber \\
   & &	:: 0.87 : 1 : 1.9 
\end{eqnarray}
which is comparable to the simple-minded approach above, and also
with the observations of lower-than-solar mass stars, although of course, 
more detailed checks with stellar models will need to be carried out to 
confirm this line of thinking. In particular, we have ignored the initial 
periods and all of the stellar evolution. 
Of course, when the surface convection zone vanishes among F stars, 
then the solar-type wind also vanishes, and angular momentum loss ceases, and
this is not incorporated here. The mass and radius dependence in Kawaler's
original formulation discussed above probably mostly accounts for this 
decrease of efficiency of angular momentum loss among higher mass stars,
as one approaches the break in the Kraft curve, and the radiative sequence
beyond. It is not adequate to describe the observations, as was 
demonstrated in the context of the Hyades by Chaboyer et al. (1995).

Thus the observed variation of $P$ with $B-V$ color 
suggests that the magnetic fields of solar-type and late-type stars on 
the {\it I} sequence, which cause angular momentum loss by coupling the 
surface of the star to the magnetized wind are also able to couple to the 
interior radiative zone, or in any event, a substantial fraction of the whole 
star. The simplest choice is that of the whole star.
The helioseismic observation of the Sun's, to zeroth order at least, 
solid-body rotation supports this interpretation. This also probably explains 
the success of solid-body models e.g. Bouvier et al. (1997) and Barnes et al.
(2001), as opposed to the more complex differentially-rotating ones, in 
accounting for a fairly large fraction of the rotation period data.
We identify this fraction with the {\it I} sequence. 
Thus the Skumanich timescale, $\tau_I$, 1\,Gyr for the Sun, is the timescale 
for the entire star to spin down, greater for a higher mass star, and smaller 
for a lower mass star, and their interiors appear to be coupled to their 
surfaces, which are in turn coupled to their winds.

\subsection{The {\it C} sequence}

Stars on the {\it C} sequence have a different behavior, whose morphology 
has been described earlier. These stars have previously also been called the 
UFRs in young open clusters.
Their sequence spins down on a different timescale, $\tau_C$. 
It is known from prior work (eg. MacGregor \& Brenner 1991; Chaboyer et al. 
1995) that angular momentum loss from these stars cannot be described by 
Kawaler's (1998) prescription. Barnes \& Sofia (1996) provided a clear 
demonstration of this impossibility despite any possible spin-up during
pre-main sequence evolution. 
For these stars, the loss rate must saturate and instead we have
$\frac{dJ}{dt} \propto - \omega$.
In practice, Chaboyer et al. (1995) and subsequent studies used a more
complex form based on modifying Kawaler's original (1988) expression,
to ensure a smooth transition between the two relationships. This form is
  \begin{equation}
  \frac{dJ}{dt}=
           { -K\Omega \, (\Omega_t^{4N/3}) (\frac{R}{R_{\odot}})^{2-N}
              (\frac{M}{M_{\odot}})^{-N/3}
              (\frac{\dot{M}}{10^{-14}})^{1-2N/3}
	   }
  \end{equation}
where $\Omega_t$ is the threshold angular velocity beyond which saturation
occurs. This might be expected if the magnetic fields saturate at high 
rotation rates, become small-scale, and fail to couple the surface 
effectively to the stellar wind. We show below that this scheme, 
although on the right track, does not correctly describe the color/mass 
dependence of the associated angular momentum loss because the mass and 
radius dependences were kept unchanged from the unsaturated case.
We also eventually show that the concept of ``saturation'' itself is 
unnecessary.

	\subsubsection{Rotational isochrones using the {\it C} sequence: }

$\frac{dJ}{dt} \propto - \omega$ implies that the rotation rates of 
these stars ought to decrease exponentially with time.
In fact, as in the case of the {\it I} sequence, we can overplot a series
of curves, again with an arbitrary color dependence which may subsequently be 
explained in terms of stellar models, over the {\it C} sequence in each 
panel. 
These form a family of curves, again parameterized by the time, $t$,
which enters exponentially, but here the color dependence and
time dependence are apparently not separable, and we had to allow the
color term, represented by the denominator in the power in the expression 
below, to change with time, 
because the bifurcation point moves redward with increasing cluster age.
The relationship we have used to generate these curves\footnote{This simple 
function is arbitrarily chosen to reflect the exponential spindown of the
sequence and the redward motion of the terminus.} is 
\begin{equation}
P = 0.2\, e ^{t/100[B-V+0.1-(1.0/3000)t]^3} 
\end{equation}
where the symbols have the same meaning as before. 
Thus these too can be associated with the previous curves to make an enhanced,
if still crude, set of rotational isochrones. These are all plotted together
in the right panel of Fig.\,4, with colors matching the respective open
cluster observations, and widths representing the fraction of cluster stars
on the {\it C} sequence. The {\it C} sequence terminates
on the blue end when the stars on it reach the {\it I} sequence. The 
timescale, $\tau_C$, for this sequence to spin down to the {\it I} sequence 
is zero for late F stars, 100\,Myr for G stars, and 200-300\,Myr for 
K-early\,M stars.

	\subsubsection{Angular momentum loss}

These stars must have a different magnetic field configuration.
At the outset, we can say that the expected small-scale magnetic fields and
the relatively mild dependence of $\frac{dJ}{dt}$ on $\omega^1$ 
(as compared to $\omega^3$ for the {\it I} sequence) suggest that the 
timescale, $\tau_C$, associated with spindown on this sequence ought to be 
very long, and therefore we ought to have $\tau_C \gg \tau_I$, which itself 
is $\simeq$ 1\,Gyr for Sun-like stars. The observations suggest otherwise.

At the risk of belaboring the point, let us re-derive the time dependence and 
retain the moment of inertia term, as before. Assuming that the relevant 
moment of inertia, $I$, is constant on the main sequence,
\begin{equation} 
\frac{dJ}{dt} \propto - \omega
\Rightarrow \frac{d(I\omega)}{dt} \propto - \omega
\Rightarrow \frac{d\omega}{\omega} = - \frac{dt}{EI}
\end{equation}
where $E$ is a constant. Integrating this expression, we get
\begin{equation}
\frac{\omega}{\omega_0} = e^{-t/EI}
\end{equation}
or 
\begin{equation}
P = P_0 \, e^{t/EI} = P_0 \, e^{t/\tau_C}
\end{equation}
so that the timescale, $\tau_C$, for spindown on this sequence is given by 
$\tau_C = EI$.

	\subsubsection{What is being braked on the {\it C} sequence? }

Now, if we assume that $E$ is relatively constant for the late-type stars
under consideration (an assumption that may need to be revisited later), 
then $\tau_C \propto I$. If $I$ represents the moment of inertia of the entire 
star, or even a fixed fraction thereof, then $\tau_C$ should increase with 
stellar mass, i.e. $\tau_C$ should increase as stars become bluer. 

The observations suggest exactly the opposite. The color dependence of the
{\it C} sequence is the reverse of that of stars on the {\it I} sequence, 
and bluer stars spin down faster.
This fact suggests that the material involved in this spindown is not that of 
the entire star but only that of the surface convection zone, since the extent 
of the convection zone has the opposite color dependence than the mass of the 
entire star. This same idea, entitled core-envelope decoupling, was first
put forward by Stauffer et al. (1984) as a way of interpreting $v \sin i$
data, but the ambiguities in these data did not succeed in convincing the
community of the correctness of this interpretation. Soderblom et al. (1993b)
developed it further in conjunction with new observations in the Pleiades,
but the idea has been somewhat marginalized by the disk-locking hypothesis 
first advanced by Edwards et al. (1993).

The data suggest that in any event,\\
$\tau_C < 800$\,Myr for F, G \& K stars since there are no {\it C} sequence 
Mt.\,Wilson stars,\\
$\tau_C \sim 300$\,Myr for early M stars (Hyades),\\
$\tau_C \sim 200$\,Myr for K stars (M\,34 \& NGC\,3532),\\
$\tau_C \sim 30$\,Myr for G stars (IC 2391/IC 2602), and\\
$\tau_C \sim 0 $ for F stars, since the {\it I} and C sequences would 
converge for these stars in the youngest clusters.

The meagerness of this timescale, coupled with the expected inefficiency of
`saturated' angular momentum loss, further bolsters the identification of the 
moment of inertia, $I$, not with that of the entire star,
which would give the opposite color dependence noted for the {\it I} sequence, 
but with that of the surface convection zone, which is known to become 
vanishingly small among the F stars, and to approach the entire stellar volume 
among the M stars. 

With some assumptions, we can do a quick calculation of the relative 
time-scales predicted by this approach as follows.
\begin{equation}
\tau_C = EI_{cz} \Rightarrow \tau_C \propto I_{cz} = \frac{I_{cz}}{I_*} I_*
\end{equation}
where $I_*$ and $I_{cz}$ are the moments of inertia of the entire star and
only the convection zone respectively.
If we use $I_* \propto (M)^{2.4}$, as before and plug in the values for 
$\frac{I_{cz}}{I_*}$ from Soderblom et al. (1993b), we get
\begin{eqnarray}
\lefteqn{\tau_{C,1.2 M_{\odot}}: \tau_{C,1 M_{\odot}}: \tau_{C,0.8 M_{\odot}} 
        : \tau_{C,0.6 M_{\odot}}: \tau_{C,0.3 M_{\odot}}   } \nonumber \\
   & & :: 0\times1.2^{2.4} : \frac{1}{14}\times1  : \frac{1}{7}\times0.8^{2.4} 
        : \frac{1}{4.7}\times0.6^{2.4}  : 1\times0.3^{2.4}   \nonumber \\
   & & :: 0 : 1 : 1.17 : 0.87 : 0.78
\end{eqnarray}
whose non-monotonicity fails to explain the M\,star observations, but to 
zeroth order is ok for the F-G-K stars.
If we use the moment of inertia values from Pinsonneault et al. (1989), again
using main sequence models at 300\,Myr, we get
\begin{eqnarray}
\lefteqn{ \tau_{C,1.2 M_{\odot}}: \tau_{C,1 M_{\odot}}: \tau_{C,0.8 M_{\odot}}
        : \tau_{C,0.6 M_{\odot}}                          }  \nonumber \\
   & & :: 0\times 10^{53.951}:\frac{1}{14}\times 10^{53.828}:\frac{1}{7}\times10^{53.656}
         : \frac{1}{4.7}\times 10^{53.428}                    \nonumber \\
   & &  :: 0\times 10^{0.951}    : 1 : 2\times10^{-0.176}
         : \frac{14}{4.7}\times 10^{-0.4}                         \nonumber \\
   & &  :: 0 : 1 : 1.33 : 1.19
\end{eqnarray}
As noted above, this does not explain the complete form of the observations 
by itself, as in the case of the {\it I} sequence, but then we have probably 
oversimplified the mass dependence, especially of saturated winds. (Note that 
we have assumed $E$ to be the same constant for all late-type stars.)
The observations show that there is an additional dependence of the angular 
momentum loss on stellar mass that will have to be decided by fitting to the 
observations. We see in a later section that $\tau_C$ increases rapidly
among lower mass stars and that $\tau_{C,0.3 M_{\odot}}$ is a few Gyr. 
Despite its obvious limitations, the simplicity of this reasoning makes it more
appealing than resorting to arbitrary, and variable, saturation thresholds.

This further suggests that the surface convection zone for stars on the 
{\it C} sequence is substantially decoupled from the inner radiative zone. 
Apparently, the magnetic field on this sequence, which fails to couple the 
surface convection zone effectively to the exterior, resulting in inefficient 
angular momentum 
loss, also fails to couple it to the interior radiative zone. We invoke 
magnetic coupling because hydrodynamic mechanisms have no particular reason
to distinguish between stars on the different sequences.
Thus, on this {\it C} sequence, stars have saturated magnetic fields, and it 
is the outer convection zone that is allowed, and observed, to spin down. If 
the star is on the {\it I} sequence, it apparently has a different magnetic 
field, solar-type, and the convective and radiative zones are connected, 
resulting in the reverse dependence of period on stellar mass or color.

Furthermore, the bifurcation point between the {\it I} and {\it C} sequences 
also moves redward with cluster age, starting at late F stars for the youngest 
clusters, and moving towards M stars with increasing age. 
Note that the break in the Kraft (1967) curve for field stars also occurs at 
late F spectral type, as does the break observable in the panels for the Mt. 
Wilson stars.
These facts suggest that a separate {\it C} sequence exists only for stars 
with surface convection zones, and that this convection zone is decoupled from
the radiative zone when the magnetic configuration is different. 
This field apparently cannot couple to the 
interior, or indeed very effectively to the exterior, to spin a star down.
The fact that the {\it C} sequence does not cross or extend past the {\it I} 
sequence suggests that the {\it C} field changes to the {\it I} field, or
becomes dominated by it, on a timescale that varies from zero for F\,stars to 
$\sim$200-300\,Myr for K-M\,stars.

Thus the difference between the two branches suggests that the two kinds of 
stars possess different magnetic field configurations. We began the discussion
of {\it C} sequence stars by invoking magnetic `saturation' but we have
shown that this is merely a way of describing the underlying convective or
turbulent magnetic field. This might provide a means of eliminating the 
arbitrary saturation thresholds that have been used in rotational modelling.
At any given moment, more than half the non-{\it I} sequence stars are on the 
{\it C} sequence, remaining on it (presumably with this small-scale magnetic 
field dominating) as they spin down to join their compatriots already on the 
{\it I} sequence. Presumably these {\it C} sequence stars contain 
faster-rotating cores. But many stars apparently jump the gun on this 
spindown, and cross the gap between the two sequences on their own.

\subsection{The gap}

We call the cuneiform region between the {\it I} and {\it C} 
sequences the `gap.' Despite the name, this region is not empty, and contains 
comparable, if smaller, numbers of stars to those on the {\it C} sequence.

The presence of stars in the gap between the {\it C} and {\it I} sequences 
suggests that some stars jump off the {\it C} sequence and migrate to the 
{\it I} sequence ahead of some of their siblings. 
This suggests that the timescale for 
the associated spindown, $\tau_G$, is therefore shorter than $\tau_C$. 
We interpret stars in the gap as those that are switching or have switched 
magnetic configuration from {\it C} to {\it I} type. This field is 
attempting to spin the star down rapidly, but is prevented from doing so on an 
extremely short timescale by the new coupling it is simultaneously attempting 
with the now fast-spinning interior.

If the field has switched, then again we have
$\frac{dJ}{dt} \propto -\omega^3$, and
$P \propto \sqrt{t/I_G}$ as on the {\it I} sequence.
The observations show that this timescale, $\tau_G$, is shorter than $\tau_C$,
which is itself much shorter than $\tau_I$, the spindown timescale on the 
{\it I} sequence, which further suggests that the moment of inertia, $I_G$, 
involved here is not that of the whole star but of a fraction of it. 
Thus, the concept of the magnetic field switching configurations and now 
attempting to couple to the (presumably faster-spinning) interior. 
The establishment of this connection is one of the crucial points we wish to
make in this paper.
The stars in the gap provide us with the evidence that the cores and envelopes
of the decoupled stars {\it (re)couple}, and tell 
us the timescale on which this coupling occurs.

We note further that if $\tau_G \approx \tau_C$, then the gap should contain 
roughly a fraction $1/e$ of the stars on the {\it C} sequence. In fact, this 
fraction appears to be greater, suggesting that $\tau_G < \tau_C$ for that 
spectral type, bolstering the interpretation above. However, it cannot be
drastically smaller, or else no stars would be observed to lie in the gap.
The observations seem to suggest that all stars pass through the gap, but
at slightly different times that reflect star-to-star variations in the
cluster.

We will have more to say about these stars in section 6 below, but we would
immediately like to note that the gap discussed here is different from the 
Vaughan-Preston gap. The two branches of stars creating that gap 
(Vaughan \& Preston, 1980) are both on the {\it I} sequence, and so the 
distinct branches of chromospheric emission in that diagram probably represent 
the existence of different kinds of dynamos for stars on the {\it I} sequence 
itself. This is a separate issue.
For a very interesting discussion of these branches, named `active' and
`inactive,' in relation to different kinds of cyclic dynamos, see Brandenburg,
Saar \& Turpin (1998) and Saar \& Brandenburg (1999). Apparently, beyond
ages of approximately 1\,Gyr, stars can have cycles on one or both branches, 
with one or the other favored by higher or lower mass stars. The latter 
publication suggests the existence of a third, `superactive' branch, populated 
by some RS\,CVn's, some spotted dwarfs, and interestingly, M\,dwarfs. A study 
of open cluster stars along these lines, suggested by the Saar and Brandenburg 
work might prove to be quite fruitful. 

Thus, we interpret the rotational morphology of solar- and late-type stars 
primarily in magnetic terms. Here, on the main sequence, we see no necessity 
to invoke disk-interaction, which might be useful, and indeed necessary, on 
the pre-main sequence. We hereby repudiate the notion of using disks to create 
the bifurcation in rotational morphology among main sequence open cluster 
stars, as we had suggested before in Barnes et al. (2001). 

However, disks might still play a role on the pre-main sequence, where the
observations suggest lower spin up than contraction would suggest.
Note that the IC\,2391/IC\,2602 observations display slower rotation 
than this simple model suggests, and that this could be attributed to disk 
interaction on the pre-main sequence.
We merely displace disks from the prime directors of the action to a
secondary place. We also believe that they affect all stars equally,
partly because of the conclusions of Barnes (2001).
We will have more to say about disks below, when we deal with pre-main
sequence stars.


\section{Connection to observations of pre-main sequence stars}

Observations of pre-main sequence stars suggest an extension of 
these ideas to even younger objects. Those of the Orion Nebula Cluster (ONC) 
by Attridge and Herbst (1992), Choi and Herbst (1996) and most recently, 
Herbst et al. (2001), including a sample of some 400 stars, suggest a 
bi-modality in the rotation periods of T Tauri stars more massive than 
$0.25 M_{\odot}$, a fact which suggests to us that the morphological 
bifurcation proposed in this paper is set at an even earlier stage on the 
pre-main sequence, and probably soon after birth. 

	\subsection{Bi-modality for $ M > 0.25 M_{\odot}$}

Following the scheme suggested by Edwards et al. (1993), Herbst and 
collaborators have interpreted the bi-modality they observe as an effect of 
magnetic interaction between the young star and its circumstellar disk. 
If this is true, then there ought to be a correlation between rotation rate 
and some indicator of disk presence or extent. Neither Stassun et al. (1999) 
nor Rebull (2001) were able to find the predicted correlation.

The possibility of using disk-interaction on the pre-main sequence to create
a wide range in initial rotation rates has prompted the construction of models
by several researchers (including the present author) that exploit this 
phenomenon to explain the rotational dispersion observed in young open clusters
(eg. Cameron \& Campbell 1993; Cameron, Campbell \& Quaintrell, 1995;
Keppens, MacGregor \& Charbonneau, 1995; Bouvier, Forestini \& Allain, 1997). 
Recent models suggest plausible disk-locking timescales of 1-3 Myr 
(Krishnamurthi et al. 1997; Sills et al. 2000; Barnes et al. 2001).
However, this paradigm appears to be refuted by the Stassun et al. (1999)
and Rebull (2001) observations. 
Furthermore, if this paradigm were correct, then the rotation rates of the
host stars of extrasolar planets ought to be lower than average when compared
to, say, the Mt.\,Wilson stars. Barnes (2001) showed that this is not the case.
These, and other problems, suggest to us that the disk interaction based
interpretation of rotational bi-modality is erroneous.

Instead, we suggest that the bi-modality observed by Herbst and collaborators
represents an early glimpse, perhaps even the first incarnation, of the {\it I} 
and {\it C} sequences observed in open clusters. Herbst et al. (2001) note 
that ``the basic features of the angular velocity distribution with mass in the
Pleiades ........ are rather precisely mimicked in our data on the ONC.''
In fact, the observations of Herbst et al. (2001) - see especially Fig.\,1 
there - look suspiciously similar to one of the panels in Fig.\,1 here. 
However, the claim of rotational bi-modality on the pre-main sequence is not 
unchallenged. Let us examine this issue in more detail, noting that the 
rotational peaks in the Herbst et al. (2001) data are at 2 and 8\,days.

In NGC\,2264, an older  pre-main sequence cluster for which a large sample of 
periods has been derived by Lamm et al. (2002), a distribution similarly 
bi-modal to Herbst's, with peaks at 1 and 4\,days, has been proposed.
Since this cluster is older than Orion, a spin-up by a factor of two may
reasonably be attributed to stellar contraction on the pre-main sequence,
modulo some disk interaction, but for all stars.
Viewed in this context, the period observations of Adams, Walter and 
Wolk (1998) of a relatively small number of stars in the Upper Scorpius OB 
Association looks suspiciously similar, and appears to detect the first 
peak at 2.5\,days, despite the fact that this conclusion does not appear to
be independently statistically supportable. The second peak is beyond the
sensitivity of their 11 night observing run.

Stassun et al. (1999), based on $\sim250$ rotation periods obtained in 
Orion OBIc/d, denied the Herbst group claims of bi-modality on statistical 
grounds, but they appear not to have checked for its presence or absence
against stellar mass. The present author believes that there is an underlying
bimodality, despite the statistical arguments made to the contrary. We note 
that here, the positions of the peaks (see Fig.\,13 of their paper) are at
1.5 and 5.5\,days. Modulo issues related to period detection efficiency -
their efficiency drops precipitously past 9\,day periods - their peak locations
suggest some spin-up relative to the Herbst et al. (2001) sample, but less than
that of the Lamm et al. (2002) sample.

Based on a similar study including nearly 300 periods 
obtained in fields flanking the ONC, Rebull (2001) denied the Herbst group 
claims of bi-modality/uni-modality for masses greater/smaller than 
$0.25 M_{\odot}$ based on statistical considerations, but the present author 
finds the claims of Herbst and collaborators believable and even convincing, 
based on his reading of the rotation period distribution displayed in Fig.\,23 
in Rebull (2001), the fact that the peaks of the distributions are located 
at suspiciously similar locations (1 and 6\,days), the fact that 
this large sample (and indeed the one of Stassun et al. 1999) might not be 
entirely coeval, the rapidity of evolution on the pre-main sequence, and the 
overall context of the main sequence observations considered earlier.
In fact, given the magnitude of these surveys, they might be even be observing 
somewhat different formation times in the ONC, as suggested for instance, by 
Dolan \& Mathieu (2002).

Thus, we believe that the existence of bi-modality in the rotation period 
observations is the key issue in this area, and that the specific locations
of the peaks can be explained mostly in terms of rotational evolution between
samples of differing ages, and secondarily in terms of observational details
of period sensitivity and detectability. It would appear that the denial of
bi-modality, admittedly statistically supported, arose from the failure to
find any correlation between rotation and disks, which were previously touted 
as the cause of the bi-modality.

	\subsection{Unimodality for $ M < 0.25 M_{\odot}$}

Herbst et al. (2001) note that the bi-modality of rotation periods in Orion 
terminates at $0.25 M_{\odot}$\footnote{The present author believes that 
bi-modality terminates at slightly higher masses on the pre-main sequence,
but this is a detail at present.}. 
Below this mass range, the distribution is 
decidedly unimodal. Thus, it appears that the {\it I} sequence does not 
continue into the mass range of the lowest mass stars, but terminates at or 
near $0.25 M_{\odot}$. All lower mass stars down to late-M spectral type are 
on the {\it C} sequence.

Lamm et al. (2002) are able to confirm this result of bi-modality/unimodality 
for rotation periods derived in the slightly older cluster NGC\,2264. 
The present author believes that the data in Rebull (2001) also reflect this 
result, and probably also the data in Stassun et al. (1999). Further evidence 
for a similar change from bi-modality to uni-modality at the same stellar mass 
may be found in $v \sin i$ observations of the Pleiades, displayed in Terndrup 
et al. (2000) - see especially Figs. 7 \& 8. Finally, we note that the three
extremely low mass cluster stars for which periods are known - APJ0323+4853
in Alpha\,Per, with a period of 7.6\,hr (Martin \& Zapaterio-Osorio 1997),
HHJ\,409 and CFHT-PL\,8 in the Pleiades, with periods of 0.258\,d and 0.401\,d
respectively (Terndrup et al. 1999), are also apparently on the {\it C}
sequence.
Higher than $0.25 M_{\odot}$ stars can apparently lie on either sequence, 
but lower mass stars are all located on the {\it C} sequence.


\section{The interface and convective sequences, the gap, and dynamo origins}

The considerations of the foregoing section suggest that the {\it I} sequence 
terminates at or near the stellar mass where full convection sets in, and the 
radiative/convective boundary within the star vanishes.
Thus, the {\it I} sequence is associated with the existence of the 
radiative/convective interface, and may more properly be termed the 
{\it interface sequence}. By extension, it can now be associated with an 
interface dynamo, of the kind suggested for instance, by Parker (1993). 
And it is probably the dynamo itself that attaches the two zones.
The associated rotational sequence attaches smoothly to the radiative sequence
in higher mass stars, but not to the {\it C} sequence.

The {\it C} sequence abruptly bifurcates from the interface sequence, 
continues past the stellar mass range where full convection sets in, 
and must be associated with a convective or turbulent 
dynamo, discussed for example by Durney, DeYoung \& Roxburgh (1993). 
Thus it may more appropriately be termed the {\it convective sequence}.
And beyond late-M and into L spectral type, there appears to be yet another 
sequence of stars where different physical processes dominate (Mohanty et al.
2002). We suspect that this rotational sequence also attaches smoothly to the 
convective sequence.

The identification of the convective/radiative interface as being crucial
to the existence of the interface sequence offers the chance of 
making a connection between solar-type stars and the Sun. In the case of the
Sun, the tachocline at the base of the convection zone is widely believed
to be the source of the dynamo (eg. Spiegel \& Weiss 1980). 
The result here would seem to bolster the case for an interface-type dynamo, 
rather than one of the Babcock-Leighton variety (Babcock 1961; Leighton 1969). 
A discussion of the pros and cons of one kind of dynamo or the other is beyond 
the scope of this work. Following up on Parker's (1993) suggestion, interface 
dynamos have been studied in some detail and developed further, including 
numerically, in both cartesian and spherical geometries respectively by 
MacGregor \& Charbonneau (1997) and Charbonneau and MacGregor (1997). 
In the opinion of the author, these results are very promising, and could be 
highly relevant to understanding rotation on the interface sequence.

At the same time, it is also clear that this cannot be the whole story,
and that the convection zone also makes a magnetic contribution, but of
a different kind, which becomes the only kind among stars below 
$0.3 M_{\odot}$, which are fully convective, and hence cannot support
the kind of dynamo activity we see in more massive stars.
Here, the dynamo activity must be of another kind, perhaps of the $\alpha^2$ 
variety (Steenbeck \& Krause 1969; Radler et al. 1990).

It appears that the case of the Sun is one of the more complicated cases,
since both a convective and an interface field contribute to its measured
properties, resulting in a mixture of characteristics that makes 
interpretation difficult. However, the difficulties may apparently be 
alleviated by studying both higher and lower mass stars.

These ideas suggest that stars do not necessarily come equipped with 
large-scale dynamos, nor can all stars host them.

If a star has a convection zone, (whose position, extent and characteristics
are decided only by a star's mass, age and metallicity), then it will have an
associated convective or turbulent and small-scale magnetic field. This field
can attach significantly to neither the inner radiative zone nor the star's 
exterior. Thus the spindown of stars with only this field is exponential and
only the outer convection zone participates.

An interface dynamo can exist only if an interface between convective and
radiative zones exists, and this fact is again decided by the fundamental
stellar properties - mass, age and metallicity. When an interface dynamo
exists, it creates a large-scale field, attaches the surface convection zone
to both the inner radiative zone and the exterior, and dominates the rotational
evolution.

It is now necessary to answer one final question: How is the interface dynamo
created? It cannot be self-created because the associated magnetic energies are
greatly in excess of equipartition energies (Parker 1993). Therefore it must
be created by an external agent. We suggest that the de-coupling between the
convective and radiative zones that is natural in the absence of the dynamo
allows the two zones to shear immediately upon star formation and associated
contraction on the Hayashi track, and that this shear creates the first dynamo.
Thus, not only does core-envelope decoupling occur, but it is also the agent 
that creates the first dynamo, which then proceeds to couple the two zones.
The shearing will continue unabated on the pre-main sequence and main sequence,
winding up any possible fossil or convective field while a star is on the
convective sequence until a dynamo is created. 

Higher-than-solar mass stars have thinner convection zones and shear more,
and hence form dynamos earlier. Lower-than-solar mass stars generate them 
later,
and fully convective stars cannot ever create them. When the dynamo is created,
it attaches the spun down convection zone to the faster spinning radiative zone
and moves the star through the gap to the interface sequence. The timescale 
for the motion through the gap is decided by the interaction between these 
various competing forces. Once the dynamo is created, in analogy with a yo-yo 
that is initially wound up and then continues oscillating back and forth, the 
dynamo can presumably sustain itself indefinitely off the
tachocline at the base of the convection zone. Finally, we note that if this
shear is indeed the origin of stellar dynamos, then because the rotation and 
shear axes must be identical, so must the magnetic axis also necessarily be 
aligned. This is so in the case of the Sun, but observation will have to 
decide the truth or falsity of this prediction for other stars.

To summarize, stars on the convective sequence have convective magnetic fields,
and are decoupled. The shear creates an interface dynamo, which couples the
zones, and drives a star through the gap till it arrives on the interface
sequence with the onset of a well-defined interface dynamo.


\section{Low mass field stars}

The $v \sin i$ observations of field M\,dwarfs in the young disk by
Delfosse et al. (1998) can be interpreted in terms of the same paradigm
by assuming that the bifurcation point between the interface and convective
sequences has moved to or near M3 by the age of the young disk 
($\sim 3$\,Gyr). Thus every bluer star than this spectral type must lie on
the interface sequence (modulo non-coevality of the sample). Every star
redward of the spectral type at which full convection sets in must lie on the 
convective sequence, which is almost vertical between M3.5 and M5.5 or M6. 
Because of the differing age and mass dependences of the two sequences, 
the convective sequence may include slower and faster rotators than the 
interface sequence in this region, and the
dispersion in $v \sin i$ should be dramatically higher in this region.
In fact, this is exactly what the observations show - see especially
Fig.\,3a in Delfosse et al. (1998). Beyond the fully convective 
boundary, old, convective sequence stars could be slower rotators than 
interface sequence stars, a fact which might explain the upper limits
in the M3-M5.5 $v \sin i$ observations.

The observations by Delfosse et al. (1998) and Basri et al. (1996) of 
M dwarfs in the old disk and halo ($\sim 10$\,Gyr) 
- see Fig. 3b in Delfosse et al. (1998) for the graphical representation - 
can now be interpreted in terms of the same model,
except that all the stars in this mass range have spun down even further
(and the bifurcation point might have moved further redward).
Apparently Delfosse et al. (1998) and Basri et al. (1996) have observed, 
in older populations, the tail of exactly the same phenomenon noted by 
Herbst et al. (2001) and others on the pre-main sequence.

The relatively small numbers of data points, the ambiguity of $v \sin i$ 
data, and the age spread of this sample do not yet permit the separation of 
the underlying interface and convective sequences. 
Rotation periods, if derivable, could clarify the issue considerably.

Mohanty and Basri (2002) were able to extend the $v \sin i$ measurements to
field stars even further down the main sequence, from M5 to L6. From M5 to
M9 or L0, they appear to be observing the spun-down convective sequence, 
and beyond that a new sequence of rapid rotation among the L dwarfs, and 
possibly overlapping with mid to late M dwarfs on the spun-down convective 
sequence. The physical processes determining the evolution of this sequence 
are different, and have been elucidated by Mohanty et al. (2002). If, as they 
suggest, there is reduced, or no angular momentum loss in these stars, then 
the rotation rates measured will reflect the initial distribution, modified 
only by the structural evolution of these objects.


\section{A schematic}

The foregoing description of the observations may be used to create a simple
schematic, which we present in Fig.\,5 in the hope that it might be useful
in interpreting future observations, especially those of older and of lower 
mass stars. 
Because this is an extrapolation in time, and in stellar mass, of behavior 
that has been documented only in a limited range of both quantities, it 
should be treated only as the roughest guide, rather than as gospel truth.
The curves plotted in Fig.\,5 use the parameterizations of equations (1) and
(15) for the indicated ages, but the convective sequences plotted in the 3\,Gyr and 
10\,Gyr panels use smaller ages (1.3 Gyr and 1.5 Gyr respectively) 
input into equation (15) than the labels on the panels suggest, 
in order to suggest better what might be the reason behind the
spectral-type dependent dispersions seen in the M\,dwarf observations.

We display the schematic in panels with ages increasing by roughly a factor 
of 3, equivalent to 0.5\,dex. This gives us a basis for comparing most of the
observational material with a panel that is suitably close in age. It suggests
the rotational evolution of stars on both the interface (solid lines) and 
convective (dashed lines) sequences, the termination of the former, and the 
mixture of fast and slow rotators on the convective sequence at great ages. 
It appears to account for all the rotational measurements in the appropriate 
stellar mass range, including those of the three very low mass stars in 
the Alpha Per (Martin and Zapatero Osorio, 1997) and Pleiades (Terndrup et al. 
1999) open clusters. The interpretation is extremely simple. The only 
complexity arises in interpreting pre-main sequence data, which we discuss 
separately below.

An instructive way to view the pre-main sequence data is to ``rotate'' 
Fig.\,1 in Herbst et al. (2001) anti-clockwise by $\pi/2$, and compare
it with one of the panels in Fig.\,1 of this paper. If one is willing to 
squint a bit, and look past the dead insects on the windshield, one 
can identify what must be the beginnings of both the interface and convective 
sequences already discussed vis-a-vis main sequence stars. Modulo the 
difficulties of assigning stellar masses on the pre-main sequence, there is 
even a hint of the radiative sequence among higher mass stars.
Because stars evolve so rapidly on the pre-main sequence and include objects
both on the Hayashi and Henyey tracks, we are here observing stars with a 
wide range of rotational properties, which results in a very real dispersion
in the measured rotation periods which seem to span a succession of sequences 
that is a compressed prelude to the later main sequence evolution. 

Note especially the extremely slow rotators at or near the fully convective 
boundary, expected to be located at higher masses like $0.4-0.5 M_{\odot}$ 
at this early age, rather than at the $0.25 M_{\odot}$ expected on the main 
sequence. In fact, here is direct evidence that stars on the convective
sequence can be slower rotators than stars on the interface sequence, in
a sense, a preview of the rotational evolution at the age of the Young or
Old Disk. In Fig.\,6 we provide a separate schematic for the interpretation
of these pre-main sequence data.
The reader might note that the Herbst et al. (2001) stars are rotating slower 
than those in IC\,2391/IC\,2602, so we are apparently directly observing 
spin-up on the pre-main sequence between these ages\footnote{Interestingly,
this spin-up is only by a factor of 2 or so, rather less than pre-main sequence
contraction would suggest, so disks are probably doing something, but as 
Rebull et al. (2002) show, they are doing something to all stars, rather than 
to create just slow rotators.}. 
These data hold the promise of revealing many other secrets about young stars.
The rapidity of pre-main sequence evolution, and the transitions from 
convective (Hayashi) to radiative (Henyey) evolution and then on to the main
sequence will undoubtedly affect the magnetic properties of stars, and might 
add considerable complexity to the understanding of the
earliest stages of rotational evolution.


\section{Connection to other stellar observables}

The rotational picture we have presented suggests the reason for the current
difficulty in understanding the X-ray and chromospheric activity on stars,
their light element abundances, and a host of other observables that depend
on stellar rotation and magnetism. Correlating any of these quantities 
against $v \sin i$ or rotation period or color or stellar mass or age or any 
other quantity will result in large scatter unless the relevant stars are
separated before into the three categories we suggest here, namely those on 
the interface sequence, the convective sequence, and in the gap. 
Researchers have apparently been attempting to correlate the same stellar 
observables with the properties of three different kinds of stars.

Upon making this
categorization, the various relationships should simplify considerably.
In particular, we expect that stars in open clusters with elevated lithium 
abundances will be found to be on the convective sequence, and that the
X-ray emission from open cluster stars will also separate cleanly into
separate dependences for each sequence. Stars with `saturated' X-ray emission
will probably be found to be located on the convective sequence.
Similar simplification should follow vis-a-vis chromospheric emission,
or indeed, any stellar observable dependent on rotation.


\section{Conclusions }

We have proposed a simple interpretation of the rotation period data currently
available for all solar- and late-type stars. 
We began by assembling color-period diagrams already available for a series 
of open clusters, and
supplementing them with new observations in M\,34 and NGC\,3532.
Using additional information about the Sun and the Mt.\,Wilson stars, we have
been able to identify distinct sequences in these diagrams. We initially call 
the two major ones in the mass range discussed here the {\it I} and 
{\it C} sequences, later suggesting the names {\it Interface} and 
{\it Convective} sequences respectively, to represent better the underlying 
mechanism, and the rotational morphology.
Higher mass stars are known to have a different rotational morphology which 
may for the present be called the {\it radiative} sequence. This sequence 
joins smoothly to the blue end of the interface sequence.
Some stars lie in the gap between the interface and convective sequences. 

The progressive evolution of the morphology with age suggests that 
color-period diagrams, although harder to construct, could be used in similar 
ways to color-magnitude diagrams, especially for young open clusters.
 
We demonstrate that the interface sequences of all clusters can be described by
one-parameter family of curves parameterized only by cluster age. 
This sequence appears to terminate on the blue end where the surface 
convection zone vanishes among the F\,stars, 
and at the point of full convection among the M\,stars at the red end.
The convective sequences, which terminate at the interface sequences on
the blue end, and continue past the point of full convection on the red end
again form a one-parameter family of curves parameterized by cluster age,
but with a different functional form.

The Sun is at the appropriate location for its age and mass.

These facts permit the construction of a crude set of rotational isochrones
and legitimize the field of `stellar gyrochronology.' The technique is 
robust enough to permit the derivation of crude ages even 
for individual field stars. It appears to be especially promising
in view of the fact that the rotation periods of stars can be determined 
with great precision, some of them to one part in ten thousand, far exceeding
the photometric precision with which stellar magnitudes can be determined.
In practice, it is limited only by stellar surface differential rotation and 
even that limitation may be alleviated by compensating for it appropriately. 

The color dependence of the morphology allows us to identify the underlying 
mechanism and simplify the understanding of main sequence angular momentum 
loss from late-type stars.
The shape of the interface sequence is primarily set by the moment of inertia
of the entire star, while that of the convective sequence is primarily set
by the moment of inertia of the surface convection zone. These two
facts suggest that the cores and envelopes of stars on the convective sequence
are decoupled, while those of stars on the interface sequence are coupled. 
Stars appear to evolve from a decoupled state on the convective sequence 
through the gap to a coupled state on the interface sequence. 
The Sun, being a relatively old star, must be in the latter state, as 
helioseismology indicates. We thus confirm the idea of core-envelope 
decoupling and specify when it occurs and when it ends. 

Because hydrodynamic coupling mechanisms have no reason to discriminate 
between stars, and cannot arbitrarily turn on or off, we suggest that the 
coupling is hydromagnetic, rather than hydrodynamic, in nature. In 
fact, it appears that the main sequence rotational morphology of open cluster 
and field stars is intimately intertwined with stellar magnetic fields. 

The rotational morphology of the interface sequence suggests that only stars 
possessing a convective/radiative interface can ever lie on it. 
Thus, it appears to have its origins in, and to be governed by, a large scale 
interface magnetic field. This field appears to couple the convective 
envelope to both the radiative core and the exterior of the star. 
The Sun is on this rotational (and apparently magnetic) sequence. 

The rotational morphology of the convective sequence is apparently governed 
by a small scale, convective or turbulent field, appropriate for an 
age-decreasing fraction of G, K, and M\,stars, which does not couple to the 
core, and only mildly to the exterior. These stars appear not to possess
large-scale dynamos.

The remaining stars in the {\it gap} between the sequences are 
apparently in transition from being dominated by a convective field to being
dominated by the interface field. These stars inform us about when and how
core-envelope (re)coupling occurs.

These facts and other evidence suggest that a small-scale convective field 
is inherent in stars with surface convection zones, but that an interface
dynamo, providing a large scale field, needs to be created. 
The natural source for the first dynamo is the shear between the decoupled 
convective and radiative zones in a star on the convective sequence, which
generates a dynamo which in turn attempts to couple the two zones, driving a 
star into the gap, and then onto the interface sequence when coupling is 
complete. This scheme predicts that the rotational axis of the star and that 
of the large-scale magnetic field are aligned.
Higher mass stars have thinner convection zones, shear more, and 
generate an interface dynamo earlier than lower mass stars.
The interface dynamo, once created, can apparently sustain 
itself indefinitely off the tachocline at the base of the convection zone.

The observational characteristics observed on the main sequence appear to
extend to stars on the pre-main sequence,
where we claim that the same behavior has been observed, and erroneously
attributed to circumstellar disks. We identify the bi-modality observed
by Herbst and collaborators with the interface and convective sequences,
and suggest that it is present in other pre-main sequence data, despite
statistically based claims to the contrary. These pre-main sequence data
additionally allow us to locate the end of the interface sequence, and to
attribute this sequence to an interface dynamo. Lower mass stars are fully 
convective, and can only lie on the convective sequence, a fact which
explains the end of the bi-modality observed by Herbst and the uni-modality
observed among lower mass stars. These stars must have a convective/turbulent 
dynamo. 

We repudiate the idea of using disk-interaction to create the bi-modal 
rotational morphology. Although disks appear to play a role on the pre-main 
sequence, disk-locking appears to affect all stars equally. We confine the 
role of disks only to that of creating the early main sequence rotational 
distribution. Even on the pre-main sequence, the bifurcation in morphology 
appears to be fundamentally explained in magnetic terms.
The concept of magnetic saturation appears to be unnecessary, being 
another way of describing magnetic activity originating in convection.

This interpretation offers a way of unifying a wide variety of disparate 
rotational measurements, ranging from those of Kraft and earlier to the 
present day, and to connect it to already well-developed magnetic ideas
about the Sun and other stars. 
It advances wind-braking theory by specifying
what exactly it is that stellar winds brake.
The new framework also appears to extend smoothly into the higher mass regime
for field stars as originally described by Kraft (1967) and 
into the lower mass regime of M stars, both young and old. 
Observations of disk and halo M\,dwarfs appear to confirm these ideas 
in two older populations of stars, but there is some confusion here that needs
to be sorted out observationally. Mohanty and Basri (2001) have apparently 
found a new sequence of even more rapid rotation among the L dwarfs, and 
possibly overlapping with late M dwarfs. The location of this sequence is 
decided by different physical processes and its existence has interesting 
parallels to those discussed in this paper.
The L dwarf sequence is suspected to connect smoothly to the red end of the 
convective sequence, while the higher-mass stars studied by Kraft are known 
to connect smoothly to the interface sequence.

To summarize, we have presented an interpretive paradigm that appears capable
of accounting for the rotational rates of all solar- and late-type stars.
The specifics of the interpretation suggest that the main sequence rotational 
evolution of solar- and late-type stars is still explicable only in terms of 
``the connection between the convective zone, the magnetic activity ..... 
and the loss of angular momentum'' (Schatzman 1962). 

Finally, we note that the paradigm outlined in this paper has wide-ranging
consequences for the interpretation of stellar magnetism and activity, X-ray,
chromospheric, or other, light-element abundances, and all stellar phenomena 
related to rotation.



\acknowledgments

SAB would like to acknowledge the support and patience of family,
friends, and teachers. Of these he would like to mention Sabatino Sofia by 
name and acknowledge him especially for the suggestion, several years ago 
during the author's thesis work, to measure rotation periods rather
than merely calculate them.
The observations presented here, and others undertaken by the author, would 
not have been possible without generous telescope time allocations from Yale 
University, Lowell Observatory, and Cerro Tololo Inter-American Observatory.
SAB also gratefully acknowledges support from Lowell Observatory, which 
nurtured the initial stages of this work through the Lowell Fellowship, from 
the McKinney Foundation through the Univ. of Wisconsin, and from the NSF 
through AST-9986962 and AST-9731302.





\clearpage 

\begin{figure}
\plotone{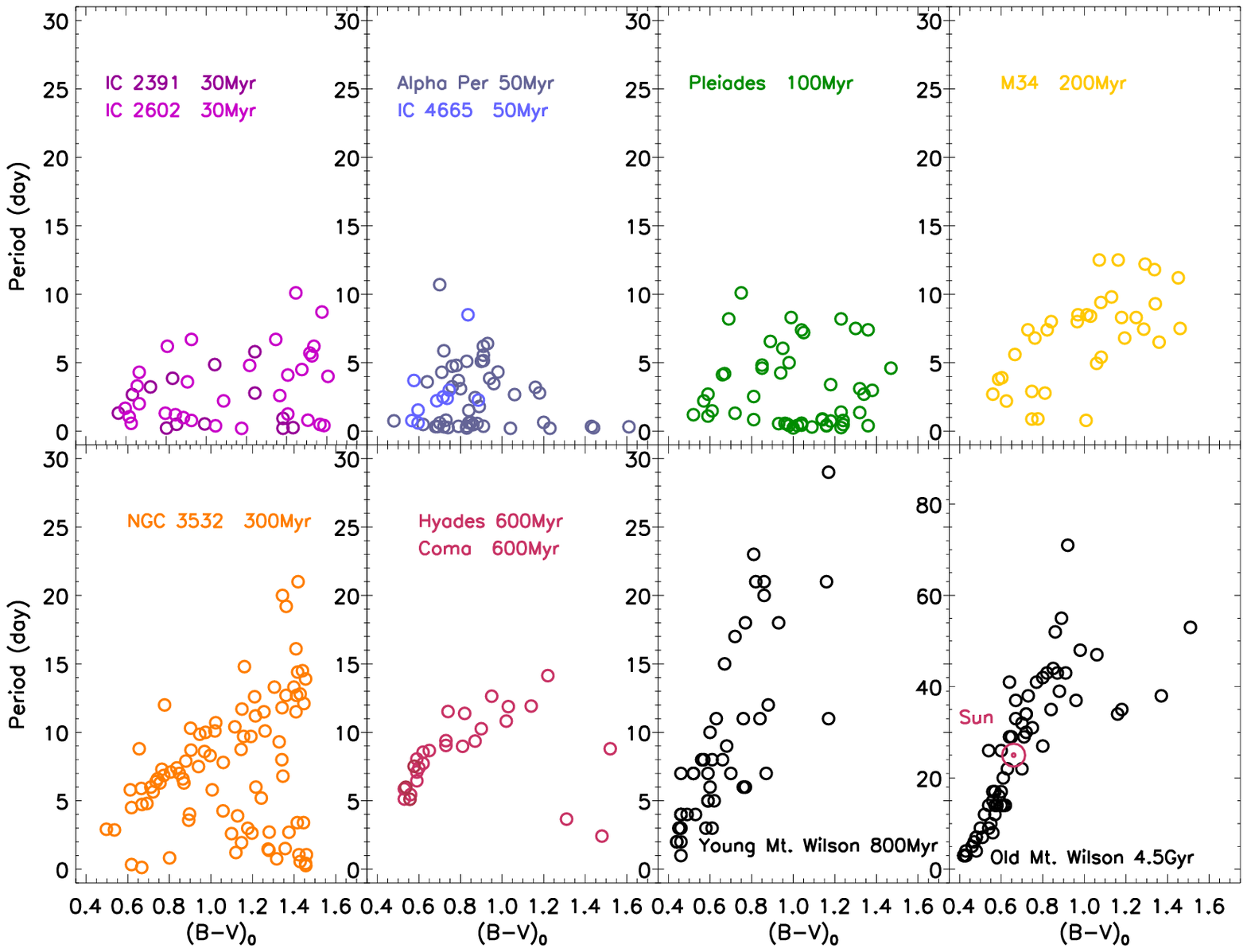}
\caption{Color-period diagrams (on a linear scale) for a series of open 
clusters and for the young and old Mt.\,Wilson stars. Note the change in
scale for the old Mt.\,Wilson stars. Two sequences of slow and fast rotators,
called I and C respectively are visible, as is a cuneiform gap between them. 
\label{fig1}}
\end{figure}

\clearpage 

\begin{figure}
\plotone{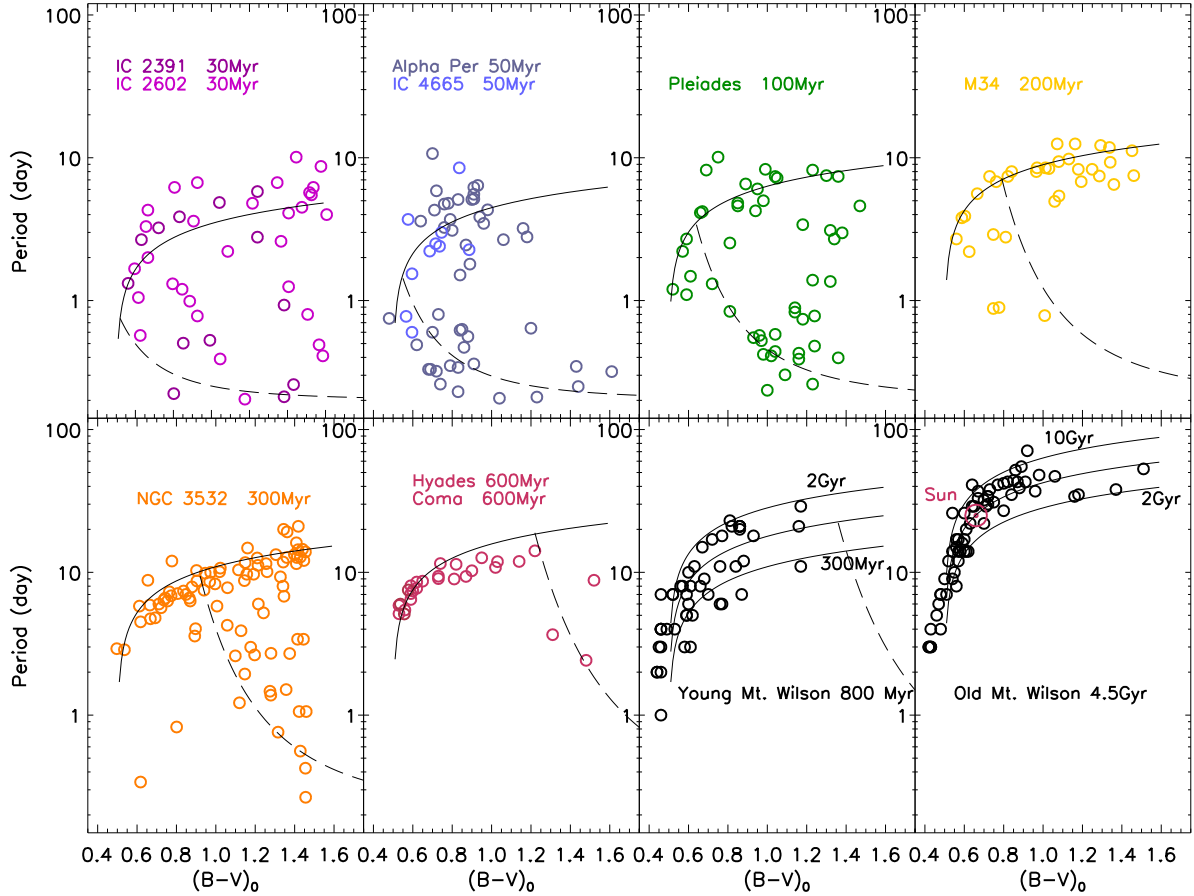}
\caption{Color-period diagrams (on a logarithmic scale) for a series of open 
clusters and for the young and old Mt.\,Wilson stars. 
The data are identical to those in Fig. 1,
and the overplotted lines, explained in the text, are meant to guide the eye 
in discerning the two sequences and the gap. These lines also constitute an 
age-parameterized family of curves.
\label{fig2}}
\end{figure}

\clearpage 
\begin{figure}
\plotone{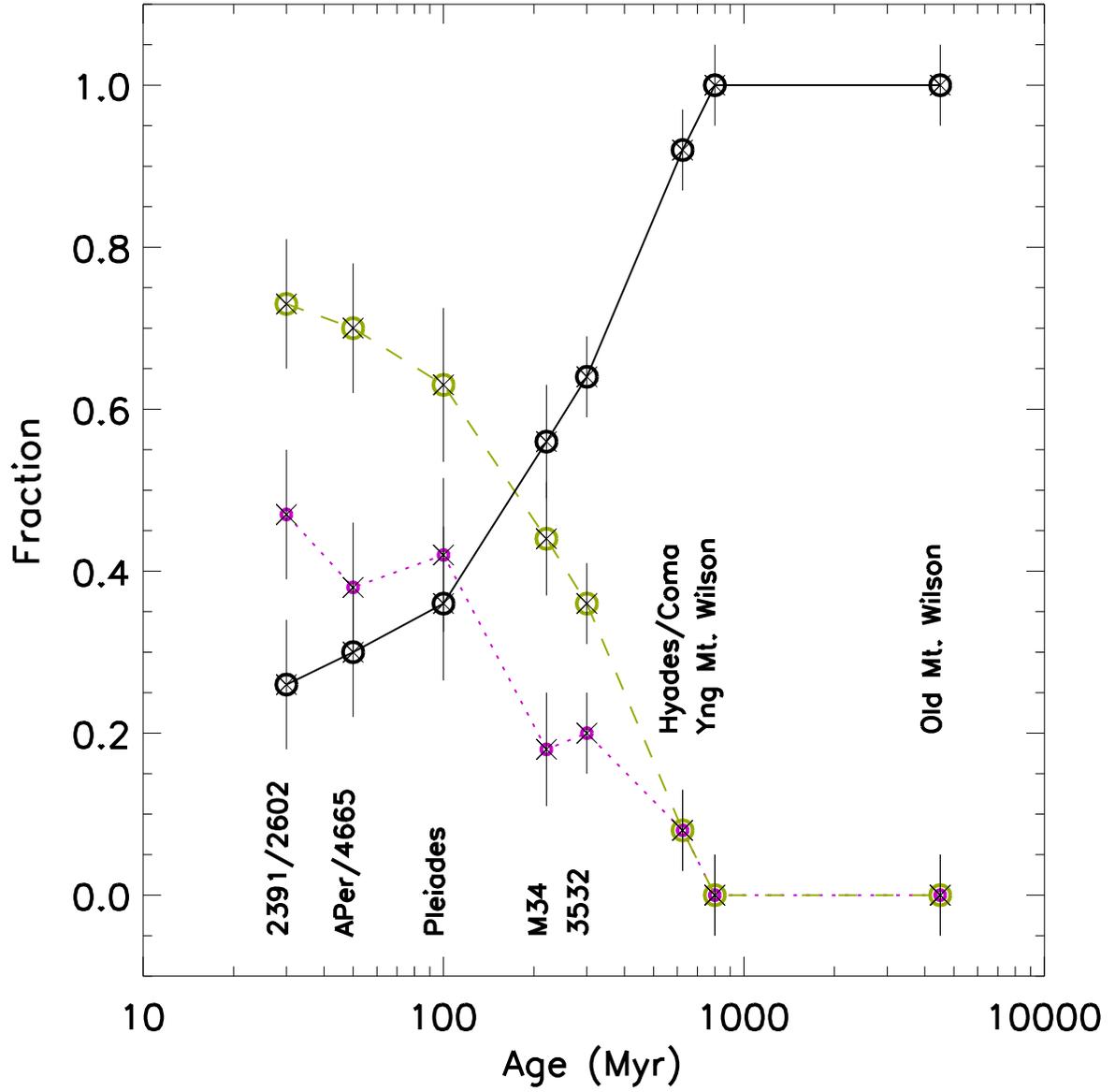}
\caption{Fraction of stars with $ 0.5 < B-V < 1.5 $ on the I
(in black) and C sequences (in violet), plotted against 
cluster age. The green symbols and line represent the sum of the 
C sequence and gap stars. Note the steady rise or decline
of the fractional numbers of each group of stars. Note that all solar-type
stars arrive on the I sequence within 800\,Myr.
\label{fig3}}
\end{figure}

\clearpage 

\begin{figure}
\plotone{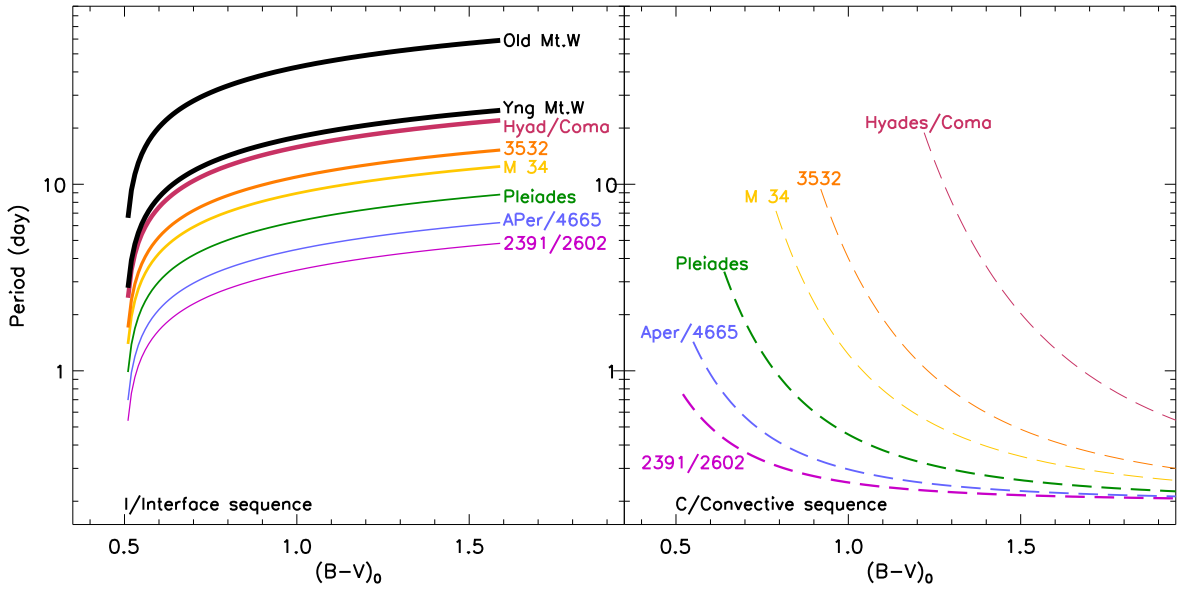}
\caption{Schematic curves representing the variation from cluster to cluster 
of the I/Interface sequence (left panel) and C/Convective sequence (right 
panel). These follow the parameterizations of equations (1) and (15) in the 
text. The thickness of each line represents the fraction of stars on the 
sequence, and colors are coordinated with data panels in previous figures. 
These constitute a crude set of rotational isochrones.
\label{fig4}}
\end{figure}

\clearpage 

\begin{figure}
\includegraphics[scale=1.5]{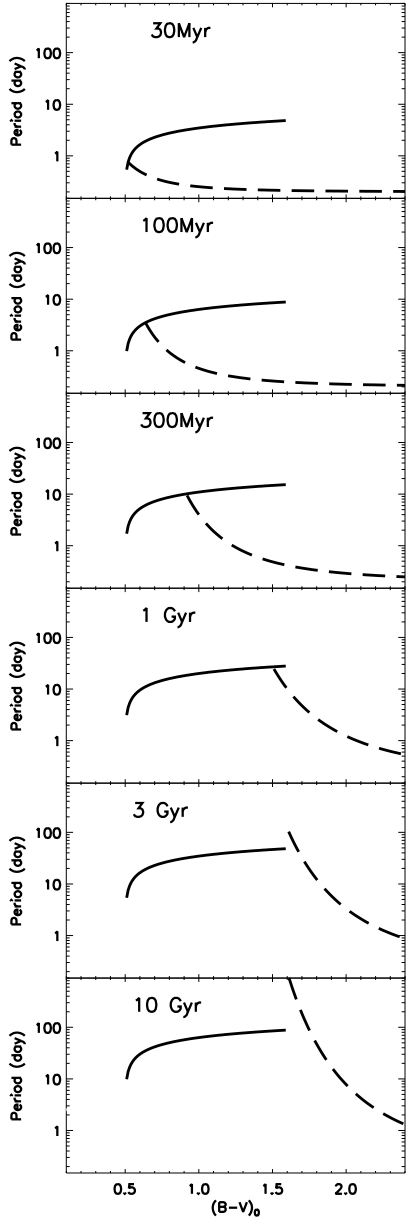}
\caption{Schematic suggesting progression of rotational evolution,
including that of late-type stars. The solid and dashed lines represent the 
interface and convective sequences respectively. 
Note especially the behavior of the M\,dwarfs. See text for plot details.
\label{fig5}}
\end{figure}

\clearpage 

\begin{figure}
\plotone{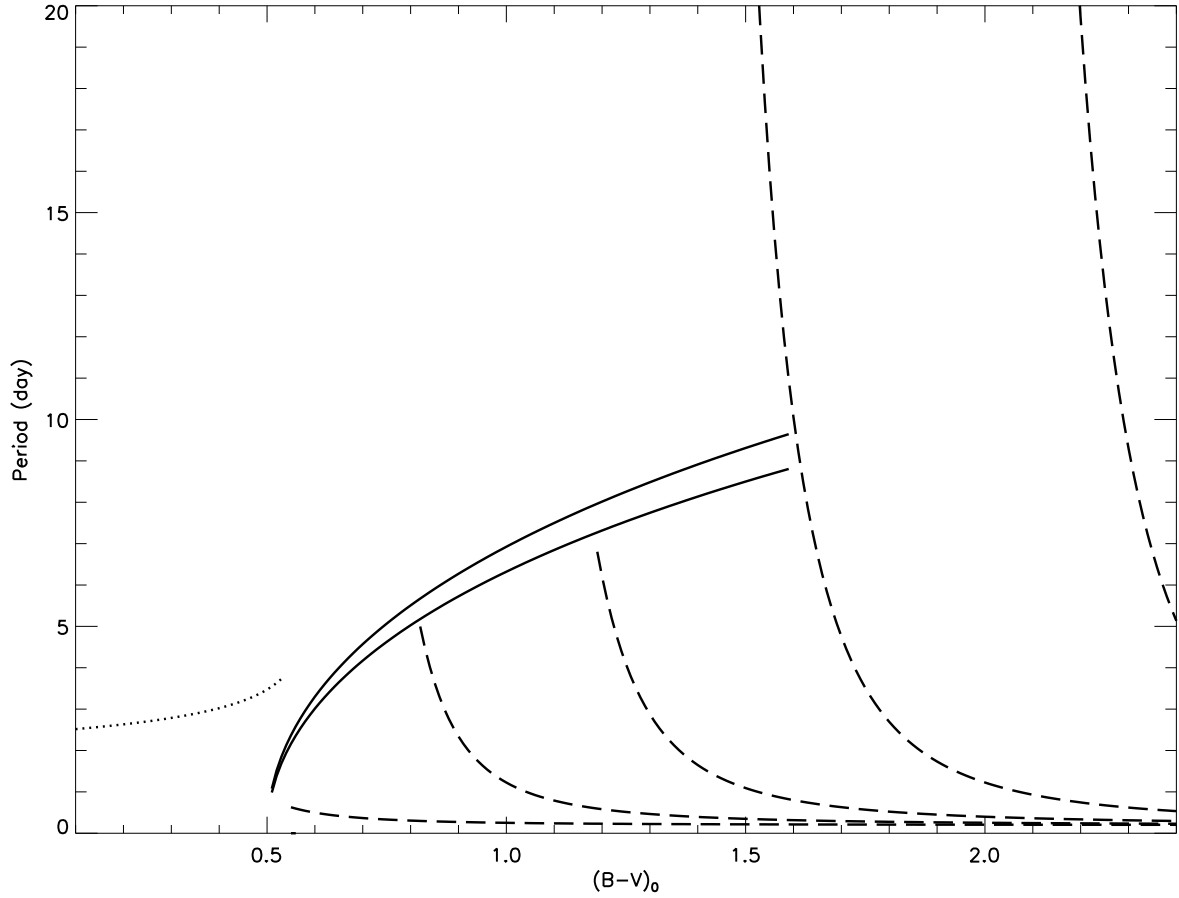}
\caption{Schematic showing interpretation of pre-main sequence data. The solid
lines represent the interface sequence, the dashed lines the convective
sequence, and the dotted line the radiative sequence. Note the termination
of the interface sequence, and the fact that fully convective stars on the
convective sequence can be extremely slow rotators.
\label{fig6}}
\end{figure}








\end{document}